\documentclass[twocolumn,showpacs,preprintnumbers,amsmath,amssymb]{revtex4}
\usepackage{bm}
\usepackage{graphicx}
\begin{document}
\title{ The Quantum Nature of a Nuclear Phase Transition.} 
\author{ 
A. Bonasera$^{a-c}$,Z. Chen$^{a}$, R. Wada$^{a}$, K. Hagel$^{a}$, J. Natowitz$^{a}$, P. Sahu$^{a}$, L. Qin$^{a}$, S. Kowalski$^{e}$, Th. Keutgen$^{d}$,T. Materna$^{a}$,T. Nakagawa$^{f}$.
}
\affiliation{
a)Cyclotron Institute, Texas A\&M, College Station, 77843, TX, USA;\\
b)Laboratori Nazionali del Sud, INFN,
via Santa Sofia, 62, 95123 Catania, Italy;\\
c)Libera Universita  Kore di Enna, 94100 Enna, Italy.\\
d)Institut de Physique Nucle\'aire and FNRS, Universite\' Catholique de Louvain, B-1348 Louvain-la-Neuve, Belgium.\\
e) Institute of Physics, Silesia University, Katowice,
Poland\\
f) Nishina center for accelerator-based science, Riken, 2-1 Hirosawa, Wako-shi,
Saitama, Japan 351-0198}

\begin{abstract}
In their ground states, atomic nuclei are quantum Fermi liquids.  At finite 
temperatures and low densities, these nuclei may undergo a phase change 
similar to, but substantially different from, a classical liquid gas phase 
transition.  As in the classical case, temperature is the control parameter 
while density and pressure are the conjugate variables.  At variance with 
the classical case, in the nucleus the difference between the proton and 
neutron concentrations acts as an additional order parameter, for which 
the symmetry potential is the conjugate variable.   Different ratios of 
the neutron to proton concentrations lead to different critical points 
for the phase transition.  This is analogous to the phase transitions 
occurring in $^{4}$He-$^{3}$He liquid mixtures.  We present experimental 
results which reveal the N/Z dependence of the phase transition and 
discuss possible implications of these observations in terms of the 
Landau Free Energy description of critical phenomena.
\end{abstract}

\maketitle

In recent times a large body of experimental evidence has been interpreted 
as  demonstrating the  occurrence of a phase transition in finite nuclei 
at temperatures  (T)  of the order of  10 MeV and at densities, $\rho$, 
less than half of the normal ground state nuclear density\cite{wci}.  
Even though strong signals for a first and a second-order phase 
transition have been found\cite{wci,bon00}, there remain a number of 
open questions regarding the Equation of State of nuclear matter near 
the critical point.  In particular the roles of Coulomb, symmetry, 
pairing and shell effects have yet to be clearly delineated.  There 
is a general consensus that in finite atomic nuclei non-equilibrium 
effects might play an important role, however,  statistical models are 
very successful in reproducing accurately a large variety of experimental 
data\cite{wci}.  This is supported by advanced experimental tecniques 
able to isolate an equilibrated region for each collision and to the 
statistical averages over those events.

A nucleus excited in a collision expands nearly adiabatically until it is 
close to the instability region  thus the expansion is 
isentropic\cite{siem}.  At the last stage of the expansion the role of 
the Coulomb force becomes very important.  In fact, without the Coulomb 
force, the system would require a much larger initial compression and/or 
temperature in order to enter the instability region and fragment. The 
Coulomb force acts as an external piston, giving to the system an 
'extra push' to finally fragment. These features are clearly seen in 
Classical Molecular Dynamics (CMD) simulations of expanding drops 
with and without a Coulomb field\cite{belk95,dorso}.  The expansion 
with Coulomb included is very slow in the later stage and nearly isothermal.  

 Even though the analogy to classical systems is quite useful, it should 
not be overemphasized as in the (T,$\rho$) region of interest,  the 
nucleus is still a strongly interacting quantum system, while at 
high T and small $\rho$ the nucleus behaves as a classical fluid.  
In particular the ratio of T to the Fermi energy at the (presumed) 
critical point is still smaller than 1 which suggests that the 
Equation of State (EOS) of a nuclear system is quite different from 
the classical one. To date this expected difference has not been well 
explored\cite{mor,wci,dago,mas,nat,ala}.   In this paper we will 
discuss experimental evidence which indicates that the phase transition 
is strongly influenced by the relative proton and neutron concentrations. 
We show that near the critical temperature for a second-order phase 
transition, the quantity I/A=(N-Z)/A behaves as an order parameter and 
the difference in chemical potential between the neutrons and protons 
is its conjugate variable. We also note that the phase transition has a 
strong resemblance to that observed in superfluid mixtures of 
liquid $^{4}$He-$^{3}$He near the $\lambda$ point.  In both systems, 
changing the concentration of one of the components of the mixture, 
changes the characteristics of the EOS. Furthermore, at high 
concentrations there is a first-order\cite{huang,land}.   The analogy 
should not be stressed too much since in our case we have two strongly 
interacting Fermi liquids while in the He mixtures we have mixed bosons 
and fermions.  However, phase transitions exhibit universal features, 
which are independent on the details of the forces and of the systems 
involved.

The experiment was performed at the Texas A\&M University Cyclotron 
Institute using the K500 Superconducting Cyclotron. Beams 
of $^{64}$Zn, $^{70}$Zn and $^{64}$Ni at 40 A MeV were incident on 
targets of $^{58}$Ni, $^{64}$Ni, $^{112}$Sn, $^{124}$Sn, $^{197}$Au 
and $^{232}$Th. Emphasis was placed upon obtaining high quality isotopic 
identification and high statistics for isotopes with Z between 3 and 16, 
for all the systems. In order to achieve this goal, a Si telescope, 
which consisted of four layers of 5cm x 5cm Si quadrant detectors of 
129, 300, 1000 and 1000 $\mu$m thicknesses was centered at  20$^{o}$. 
In this telescope four Li isotopes and six to seven isotopes of each 
element from  Z =4 up to Z = 16 were clearly identified with an energy 
threshold ranging from  5 A MeV for Li isotopes to 15 A MeV for Si 
isotopes. In addition 16 CsI light charged particle detectors and 16 
neutron detectors were also placed around the target to detect 
coincident light particles. In this paper, however, only results for 
isotopes measured inclusively in the telescope are presented. In order 
to get the angle integrated yields of isotopes, the energy spectra 
observed at 17.5$^{o}$ and 22.5$^{o}$ were fit using a moving source 
parameterization.  In the following we will show evidence that, when 
scaled properly, the yields for all the systems studied behave in a very 
similar fashion, a necessary condition for systems undergoing phase 
transitions.

The key factor of our analysis is the value $I$, proportional to the 
third component of isospin, of the detected fragments.  For instance, 
a plot of the yield versus mass number when I=1 displays a power law 
with exponent $\tau\approx 2.4$.  This is shown in fig.(1) for the 
$^{64}$Ni+$^{232}$Th case at 40 MeV/nucleon. Similar plots for I=3 give 
a much flatter distribution.  In Figure 1 we have made separate fits for 
each case sorting odd Z(open symbols) or even Z(filled symbols) separately. 
 As we see we obtain four different curves which suggests that pairing is 
playing a role in the dynamics. Notice that the plotted yields are for odd A 
nuclei for which we expect pairing to be zero\cite{pres}.  This implies that the observed 
fragments have emitted at least one neutron before cooling down.  In that 
case the parent nuclei would be even-even and odd-odd nuclei\cite{marie}.  
This could explain the shifts of the distributions, in particular with the 
even-even yields being flatter than the odd-odd cases since those fragments 
are more bound.  In general, the mass distributions for  $-2\leq I\leq 4$, 
when fitted with a  power law, give exponents ranging from 2.5 to 0.  For 
our data, using yields of fragments with I = 0 or 1 we find an average 
value of $\tau=2.3 \pm 0.1$.

\begin{figure}[ht]
\centerline{
\includegraphics[width=3.0in]{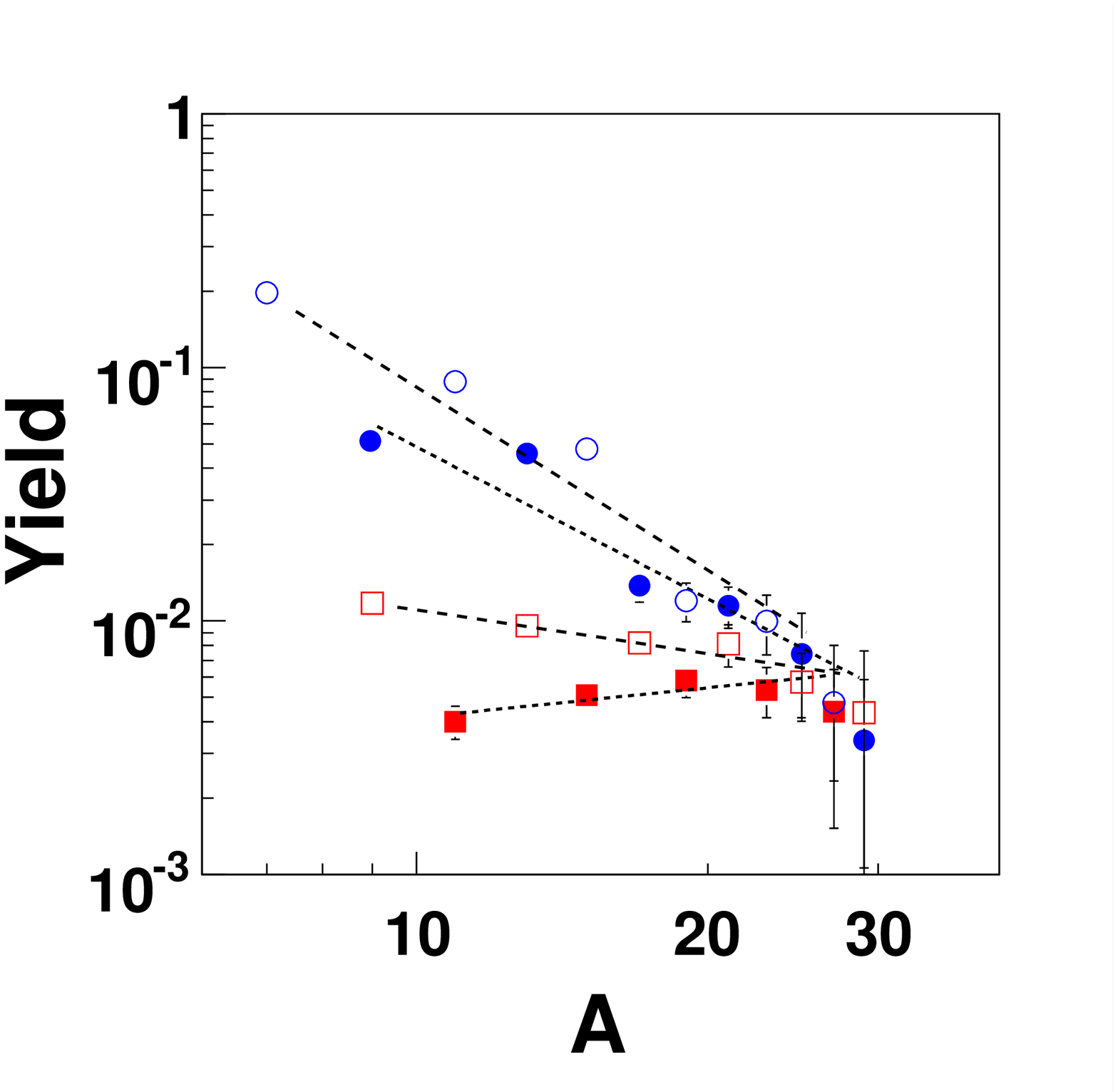}}
\caption{Mass distribution for the $^{64}$Ni+$^{232}$Th system 
at 40 MeV/nucleon for I=1 (circles) and I=3 (squares) . The 
lines are power law fits with exponents $\tau=2.40\pm0.08$, $1.99\pm0.17$, 
$0.57\pm0.11$, - $0.39\pm0.26$ respectively.  For each case the fits are 
shown separately for odd (open symbols) and even (filled symbols) detected Z.
}
\label{fig1}
\end{figure}

The observance of this power law suggests that the mass distributions may 
be discussed in terms of a modified Fisher model\cite{bon00}:  

\begin{equation}
  \label{eq:fisher}
  Y=y_0A^{-\tau}e^{-\beta \Delta \mu A},
\end{equation}
where y$_{0}$ is a normalization constant, $\beta$ is the inverse 
temperature and $\Delta\mu = F(I/A)$ is the difference in chemical 
potential  between neutrons and protons, i.e.,  the Gibbs free energy per 
particle, $F$,  near the critical point.  If we accept that $F$ is dominated 
by the symmetry energy and make the approximation that 
$F(I/A) = 25(I/A)^{2}$ MeV/A,  i.e. the symmetry energy of a nucleus in 
its ground state\cite{pres}.  We will use this relationship in order to 
infer an approximate value of the temperature of the system.  However, 
we stress that in actuality, F(I/A) is a function of density, temperature 
and all other relevant quantities near the critical point.  In the 
literature\cite{wci} ground state values of the symmetry energy 
coefficient or of binding energies of nuclei are often employed to 
derive the temperatures reached by the nuclei in the collisions.  
This is not generally correct since the values appropriate to the 
density and temperature sampled should be used.  Since the ratio of 
the chemical potential to the temperature enters equation(1) or other 
similar equations obtained from statistical models\cite{wci}, deriving 
the true value of the temperature from yields or yield ratios alone is 
possible only when the chemical potential is known.

According to the Fisher equation given above, we can compare all systems 
on the same basis by normalizing the yields and factoring out the power 
law term.  For this purpose we have chosen to normalize the yield data 
to the $^{12}$C yield, i.e. we define a ratio:

\begin{equation}
  \label{eq:R}
  R =\frac{Y A^{\tau}}{Y(^{12}C) 12^{\tau}},
\end{equation}

The choice to normalize to $^{12}$C is not arbitrary.  We want to scale to 
the power law dependence in the equation (1) and for this reason we choose 
a nucleus, which belongs to a power law distribution ($I = 0, 1$ cases 
discussed above).  Using different nuclei belonging to the same groups 
gives results similar to the ones discussed below.  On the other hand 
using nuclei from groups (e.g. I = 4) exhibiting smaller power law $\tau$ 
parameters introduces a spurious constant/A dependence that destroys the 
scaling discussed below.  The normalized ratios for the system 
$^{64}$Ni + $^{64}$Ni at 40 MeV/nucleon are plotted as a function of the 
(ground state) symmetry energy in figure (2).  The data display an 
exponential decrease with increasing symmetry energy, except for the 
isotopes for which $I = 0$.  The yields of these I = 0 isotopes are 
clearly not sensitive to the symmetry energy but rather to the Coulomb 
and pairing energies and possibly to shell effects.  The appearance of 
two exponential curves with the same exponent will become clear from the 
discussion below.  A fit using eqs.(1) and (2)  gives an 'apparent 
temperature' $\Theta$ of 6.0 MeV.  This value of $\Theta$ would be the 
real one if $\mu = 25$ MeV (the g.s. symmetry energy coefficient value) 
and if secondary decay effects are negligible.  In general we expect that 
the symmetry energy coefficient is density and temperature dependent, as 
we will discuss in the framework of the Landau free energy approach below.  
We stress that the appearance of two branches in fig. 2, clearly indicates 
that the free energy must contain an odd power term in (I/A)  at variance 
with the ground state symmetry energy.

\begin{figure}[ht]
\centerline{
\includegraphics[width=3.00in]{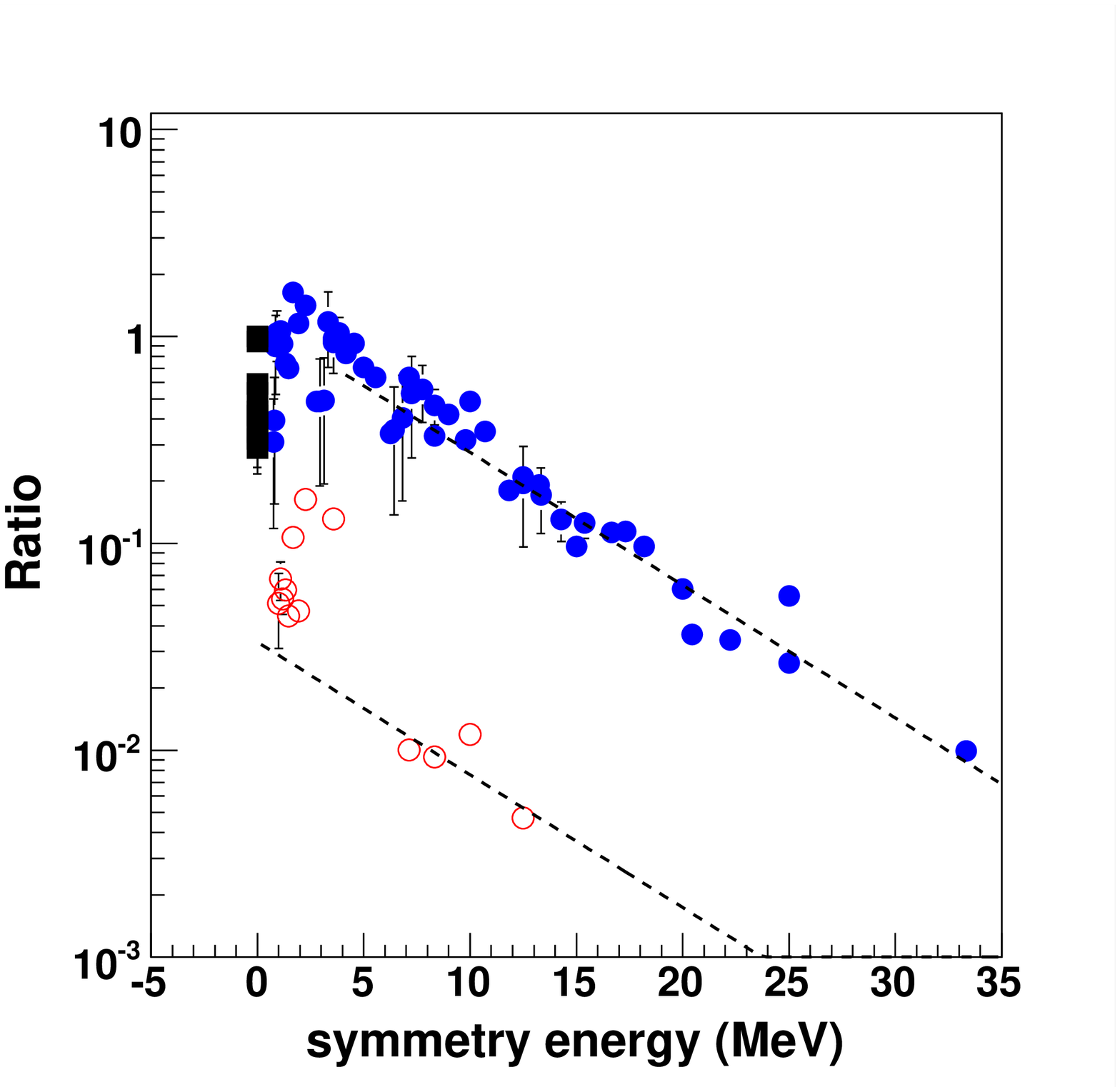}}
\caption{
Ratio versus symmetry energy for the $^{64}$Ni + $^{64}$Ni case at 40 
MeV/nucleon.  The dashed lines are fits using a ground state symmetry 
energy, eq. 1, and an apparent 'temperature' of 6 MeV. The 
$I < 0$ and $I > 0$ isotopes are indicated by the open and full circles 
respectively.  The $I = 0$ cases by the full squares. 
}
\label{fig2}
\end{figure}

To further explore the role of the relative nucleon concentrations we plot 
in figure (3) the quantity $\frac{F}{T}=-\frac{ln(R)}{A}$  versus (I/A) i.e. 
the neutron to proton concentration.  As expected the normalized yield 
ratios depend strongly on (I/A). Pursuing the question of phase transition 
we can perform a fit to these data within the Landau description. In this 
approach the ratio of the free energy to the temperature is written in 
terms of an expansion:

\begin{equation}
  \label{eq:order}
 \frac{F}{T}=\frac{1}{2}a m^2+\frac{1}{4}b m^4 +\frac{1}{6}c m^6-m\frac{H}{T}
 \end{equation}
where $m = (I/A)$ is an order parameter, $H$ is its conjugate variable and 
$a-c$ are fitting parameters.  The introduction of the conjugate variable 
$H$ is necessary since the order parameter can be obtained from the 
derivative of the Gibbs free energy with respect to H\cite{huang}.  A 
practical consequence of this is that we expect that the ratio plotted 
in fig. 2 should display two branches the lower one referring essentially 
to proton-rich and the higher one to neutron-rich isotopes.  In fig. 2 
only a few isotopes are seen in the lower yield branch since other 
proton-rich nuclei either have a short lifetime or evaporate some 
particle before reaching the detector.  Microscopic Antisymmetrized 
Molecular Dynamics (AMD)\cite{ono} calculations, which are dynamical 
calculations taking into account the Pauli principle as well as a 
realistic nuclear mean field, allow the identification of the primary 
excited fragments and clearly indicate the existence of such a lower 
branch in the first stages of the reaction. The existence of the two 
branches also clearly appears from the quantities plotted in fig. 3.

\begin{figure}[ht]
\centerline{
\includegraphics[width=2.75in]{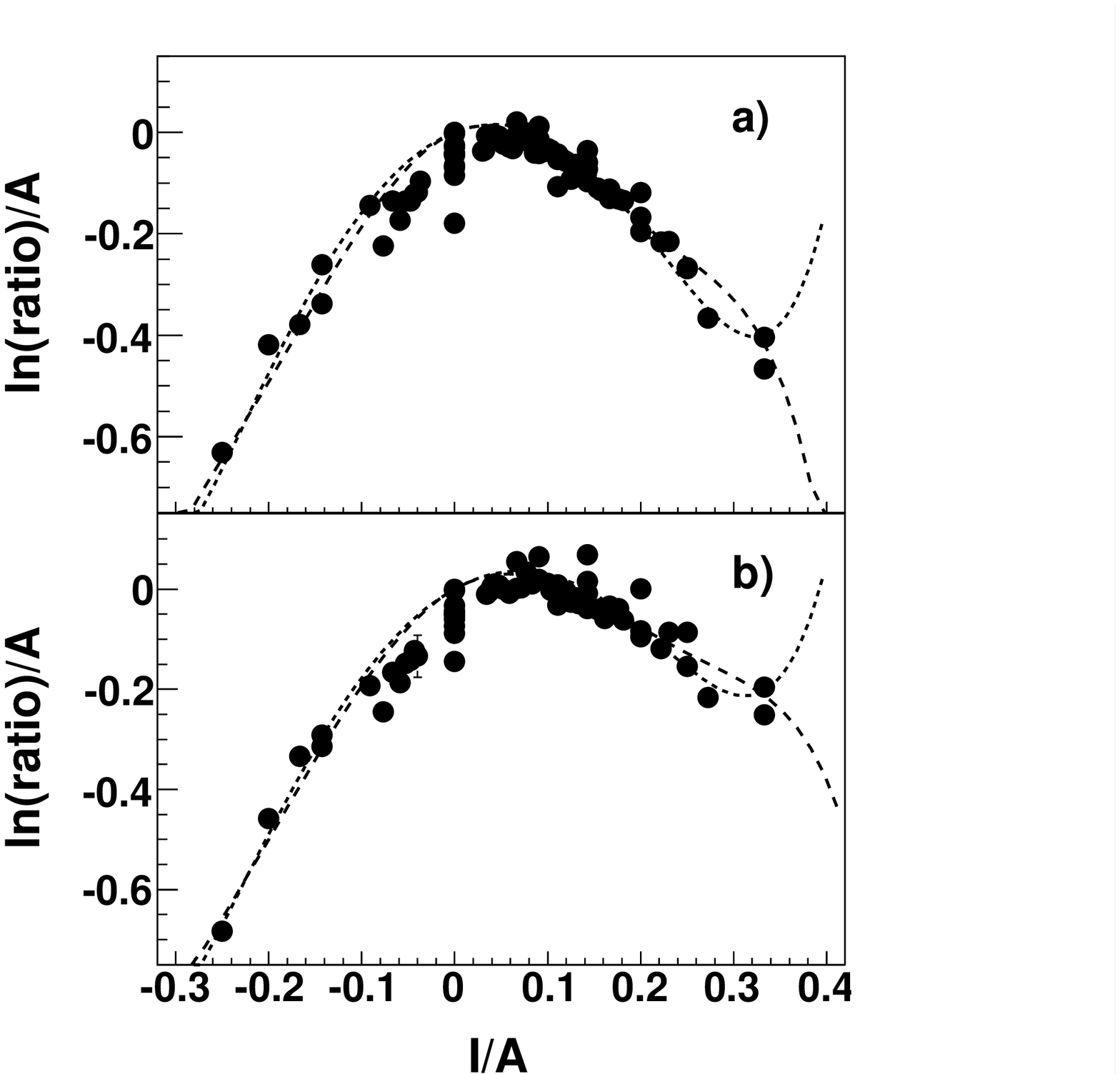}}
\caption{
(Minus) free energy versus symmetry term  for the case $^{64}$Ni+$^{58}$Ni 
(upper panel) and $^{64}$Ni+$^{232}$Th at 40 MeV/nucleon (lower panel).  
The dashed lines are fits based on Landau O($m^{6}$) free energy either 
for a second-order(long-dashed) or first-order (short-dashed) phase 
transition (see text).
}
\label{fig3}
\end{figure}

We stress that the use of the Landau approach is for guidance only and 
now proceed to discuss some possible scenarios that should be tested in 
future studies.  If we force the parameter $c$ = 0 in eq. 3, i.e. we 
reduce the Landau free energy to fourth order (which implies a 
second-order phase transition with given critical 
exponents\cite{huang,land}) and we fit the data of fig. 3 we obtain 
$a = 19.2$ and $b = -130.73$.  This result is unphysical since it implies 
that the free energy is negative for large $m$\cite{land}.  A fit using 
eq. 3 gives the following values for the $^{64}$Ni + $^{58}$Ni 
( $^{64}$Ni + $^{232}$Th) : $a = 23.5(18.86)$, $b = -413.8(-260.3)$, 
$c = 2848.3(1408.1)$ and $H/T = 0.79(1.06)$ and is displayed in 
fig. 3(long-dashed line).  This case is discussed in detail 
in \cite{huang} and corresponds to a 'classical' second-order phase 
transition.  However it also indicates that a tricritical behaviour is 
possible for different temperatures and/or densities and could be 
obtained in different experimental conditions.  There are some features 
that it is important to emphasize:

1) In Landau's approach a line of first-order phase transitions is given 
by the condition 

\begin{equation}
\label{eq:firstorder} 
b=-4\sqrt{ca/3}
\end{equation}   
Using the values of $a$ and $c$ given above we get $b =-597.5 (-376.3)$ 
which is close to the fitted values given above.  In particular if we 
substitute eq.(4) into eq.(3) and perform the fit of the experimental 
data given in fig. 3, we obtain a result (given by the short-dashed line) 
of a similar quality to that seen for the case of a 'classical' 
second-order phase transition.  In the case of a first-order transition 
the fit bends up for large (positive) $m$, however the data do not 
distinguish between the two cases.  In fact, to explore this possibility 
requires high precision data for very asymmetric isotopes even for small 
mass numbers.  This is beyond the capability of the present data.  

2) The position of the maximum in fig. 3 (minimum of the free energy) is 
displaced from I = 0, and depends on the proton/neutron ratio of the 
emitting source.  For N = Z sources we would expect the maximum at  $I = 0$.
Nevertheless a discontinuity remains since the fragments that have $I = 0$ 
have different yields.  

These features are reminiscent of a superfluid $\lambda$ transition 
observed as some $^{3}$He is added to $^{4}$He \cite{huang}.  Starting 
from pure $^{4}$He which has a critical temperature of 2.18 K, the 
critical temperature for the second-order transitions decreases with 
increasing $^{3}$He concentration until at a lower temperature,  
$T = 0.867$K, a first-order transition appears. This point is known as 
the tricritical point for this system.  Similarly, a nucleus, which should 
undergo a liquid-gas phase transition, is influenced by the different 
neutron to proton concentrations.  Thus the discontinuity observed in 
fig. 3 could be a signature for a tricritical point as in the 
$^{4}$He-$^{3}$He case.  We believe that our data, together with the 
use of the Landau O($m^{6}$) free energy, suggest such a feature but 
are not sufficient to clearly demonstrate it.  Some other 
works\cite{campi,gulm}, also suggest that a line of critical points 
might be found away from its 'canonical' position, i.e. at the end of 
a first-order phase transition and, for small systems, even extending 
into the coexistence region.    

3) Our choice of the free energy given by eqs. 3 and 4, if analytically 
extended (H/T = 0) for large asymmetries displays three equal minima.   
However we would like to stress that such minima might not be found 
experimentally, because secondary evaporation effects can strongly 
influence the yields of isotopes far from symmetry.

4) As expected the position of the maximum, i.e. of the critical value of 
$(I/A)_{c}$ depends on the $(I/A)$ of the source.  This is shown in fig. 4 
where such a critical value is plotted as a function of the symmetry 
divided by the Coulomb energy of the compound system.  We note this 
dependence on the compound nucleus asymmetry but recognize that the actual 
value of the peak could be shifted because of secondary evaporation 
effects.  We have explored this using the AMD code of A. Ono\cite{ono} 
to determine yields of both the primary (300 fm/c) hot fragments and 
secondary cold fragments remaining after fragment de-excitation.  In 
fig. 4 we display these calculated yields. As we see from the calculations 
there is indeed a shift even though the qualitative features of the 
phase transition are preserved.  Notice that in the calculations the 
critical value $(I/A)$ is not zero even for collisions of initially 
symmetric N = Z nuclei.  However, we would predict that experimental 
data for such a system would exactly give zero since the calculations 
are systematically higher than data for the systems studied here.  We 
stress that if the critical $m = (I/A)$  is zero (i.e. $H = 0$)  it 
implies that the order parameter is zero above the critical 
point\cite{huang,land}.  In the cases studied here, since $H\neq 0$ in all 
cases, in agreement with the results of phase transitions in presence of an 
external field\cite{land}, we suggest that the order parameter $m$ is never 
zero even above the critical temperature.  This implies that above the 
critical point the system has a lower degree of symmetry than in the case 
with  $H = 0$.  An experimental investigation of this feature would be very 
interesting and this could be accomplished in collisions of  N = Z nuclei 
(eg. $^{40}$Ca + $^{40}$Ca) as function of beam energy in the same 
conditions as the present measurements.

\begin{figure}[ht]
\centerline{
\includegraphics[width=2.75in]{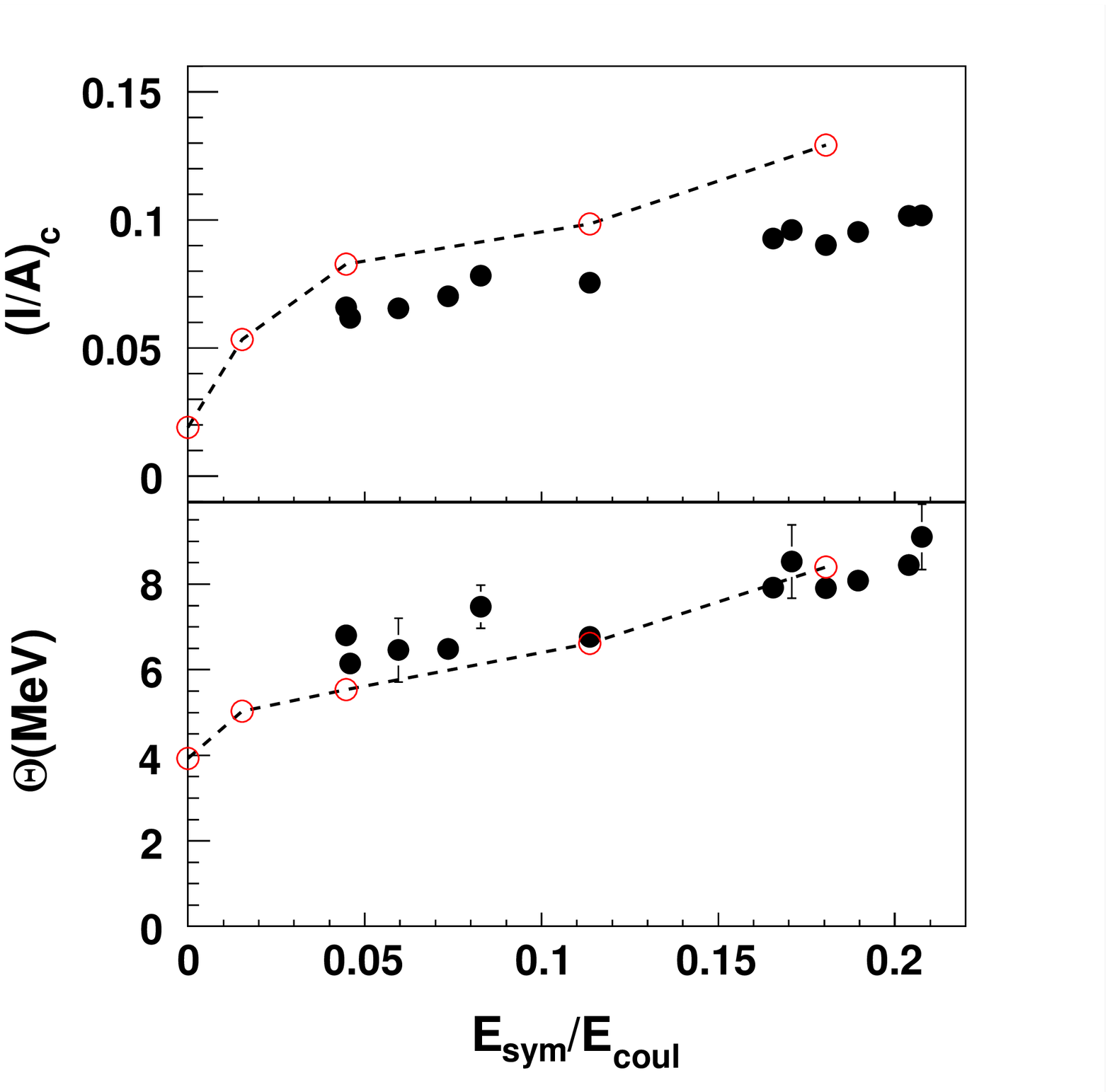}}
\caption{
'Critical' I/A (upper panel solid circles) and 'apparent temperature' 
(lower panel solid circles) versus ratio of the compound nucleus symmetry 
to Coulomb energies for all the experimentally investigated systems.  
The same quantities are plotted for AMD calculations (open symbols). The 
lines are drawn to guide the eye.
}
\label{fig4}
\end{figure}

In the same fig. 4 we plot the apparent temperatures 'T' obtained as in  
fig. 2 for all the studied systems.  As we can see 'T' increases with 
increasing asymmetry.  To understand this result we note that increasing 
asymmetry is associated with increasing mass numbers of the source.  Thus 
we could expect the critical 'T' to increase because of a reduction of 
the surface term with respect to the volume term. However, at the same 
time we increase the Coulomb and the symmetry term contributions which 
would normally act to reduce the critical temperature.  Because the 
system expands, it is apparently able to rearrange in such a way as to 
reduce the effect of the Coulomb force as much as possible.  For this 
reason we have chosen in fig. 4 to plot the displayed quantities as a 
function of the ratio of the (g.s.) symmetry energy to the corresponding 
Coulomb energy.  We expect that when this ratio approaches 1 the 'T' 
value should start to saturate and eventually decrease.

In conclusion, in this paper we have presented and discussed experimental 
evidence for the observation of a quantum phase transition in nuclei, 
driven by the neutron/proton asymmetry. Using the Landau approach, we 
have derived the free energy for our systems and found that it is 
consistent with a line of first-order phase transitions terminating at 
a point where the system undergoes a second-order transition. The 
properties of the critical point depend on the symmetry and the 
critical temperature increases for increasing asymmetry (and source size).  
This is analogous to the well known superfluid  $\lambda$ transition 
in $^{3}$He-$^{4}$He mixtures. We suggest that a tricritical point, 
observed in $^{3}$He-$^{4}$He systems may also be observable in 
fragmenting nuclei.  These features call for further vigorous experimental 
investigations using high performance detector systems with excellent 
isotopic identification capabilities. Extension of these investigations 
to much larger asymmetries should be feasible as more exotic radioactive 
beams become available in the appropriate energy range.  

Exploration of quantum phase transitions in nuclei is important to our 
understanding of the nuclear equation of state and can have a significant 
impact   in nuclear astrophysics, helping to clarify the evolution of 
massive stars, supernovae explosions and neutron star formation. 

One of us (A.B.) thanks the Cyclotron Institute at Texas A \& M University 
for the warm hospitality and support. AMD calculations have been performed 
on the Riken Super Computer Center and we acknowledge Dr. H. Sakurai for 
making this facility available to us.  This work was supported by the 
United States Department of Energy under Grant \# DE-FG03- 93ER40773 and 
by The Robert A. Welch Foundation under Grant \# A0330.

\end{document}